\documentclass[proceedings]{JHEP3}

\PrHEP{PrHEP hep2001}                   % Do not change this one
\conference{International Europhysics Conference on HEP}                % Do not change this one

\usepackage{epsfig}			% please use epsfig.

\newcommand{\beq}{\begin{equation}}
\newcommand{\eeq}{\end{equation}}
\newcommand{\bea}{\begin{eqnarray}}
\newcommand{\eea}{\end{eqnarray}}

\newcommand\sss{\scriptscriptstyle}
\newcommand\as{\alpha_{\sss S}} 
\newcommand\aem{\alpha_{\rm em}}

\def\epem{e^+e^-}
\def\ord{{\cal O} }
\def\MS{$\overline{\rm MS}$}
\def\bentarrow{\:\raisebox{1.3ex}{\rlap{$\vert$}}\!\longrightarrow}

\newbox\mybox
\newcommand\fverb{\setbox\mybox=\hbox\bgroup\verb}
\newcommand\fverbdo{\egroup\medskip\noindent\fbox{\unhbox\mybox}\ }
\newcommand\fverbit{\egroup\item[\fbox{\unhbox\mybox}]}

\font\beeg=cmr17 scaled 1600		% Stylish initials
\newcommand\init[1]{\setbox\mybox=\hbox{{\beeg #1}~}%
		   \noindent\global\hangindent=\wd\mybox\global\hangafter-2%
		   \sc\smash{\llap {\lower 13.2pt \box\mybox}}}

\title{QCD radiative corrections to $\gamma^*\gamma^* \to$ 
hadrons at LEP2}

\author{M. Cacciari\\ 
Dipartimento di Fisica, Universita' di Parma\\ and
INFN, Sezione di Milano, Gruppo Collegato di Parma\\
        E-mail: \email{cacciari@pr.infn.it}}

\author{\speaker{V. Del Duca}\\
	I.N.F.N., Sez. di Torino, via P. Giuria, 1 - 10125 Torino, Italy\\
        E-mail: \email{delduca@to.infn.it}}

\author{S. Frixione\thanks{On leave from INFN, Sez. di Genova, Italy}\\
LAPP, Chemin de Bellevue, BP 110,
74941 Annecy-le-Vieux CEDEX - France\\
	E-mail: \email{Stefano.Frixione@cern.ch}}

\author{Z. Tr\'ocs\'anyi\\
Univ. of Debrecen and
Inst. of Nuclear Research of the Hungarian Academy of Sciences\\ 
H-4001 Debrecen, PO Box 51, Hungary\\
        E-mail: \email{zoltan@zorro.atomki.hu}}

\abstract{In this talk we describe the order-$\as$ corrections to the total 
cross section and to jet rates in $\gamma^* \gamma^* \to$~hadrons
for the process $\epem\to\epem +$~hadrons.
We use a next-to-leading order general-purpose partonic Monte Carlo 
event generator that allows the computation of a rate differential 
in the produced leptons and hadrons. We compare our results with the 
experimental data for $\epem\to\epem +$~hadrons at LEP2.
}

\begin{document} 

Strong interaction processes, characterised by a large kinematic scale,
are described in perturbative QCD by a fixed-order expansion in $\as$
of the parton cross section, complemented, if the scattering process is
initiated by strong interacting partons, with the Altarelli-Parisi evolution
of the parton densities. In many scattering processes, the-state-of-the-art
computation of production rates is at the next-to-leading order (NLO). 
However, in kinematic regions characterised by two large and disparate 
scales, a fixed-order expansion may not suffice: large logarithms of the
ratio of the kinematic scales appear, which may have to be resummed.
In processes where the centre-of-mass energy $S$ is much larger than the
typical transverse scale $Q^2$, the leading logarithms of type $\ln(S/Q^2)$
are resummed by the BFKL equation.
Several observables, like the scaling
violations of the $F_2$ structure function,
forward-jet production in DIS,
dijet production at large rapidity intervals
and $\gamma^* \gamma^* \to$ hadrons in $\epem$ collisions have been 
measured~\cite{Acciarri:1999ix,lin,Abbiendi:2001tv,Prange} and 
analysed~\cite{Bartels:1996ke,Cacciari:2001cb} in this fashion.

In this talk we report on a NLO calculation~\cite{Cacciari:2001cb} of the 
total cross section and
of jet rates in $\gamma^* \gamma^* \to$ hadrons for the process
$\epem\to\epem +$~hadrons at LEP2, and we compare the NLO calculation to 
the data from the CERN 
L3~\cite{Acciarri:1999ix,lin}, OPAL~\cite{Abbiendi:2001tv} 
and ALEPH~\cite{Prange} Collaborations. Namely, we consider
\beq
\begin{array}{rcl}
e^+ + e^- & \longrightarrow & e^+ + e^- + \underbrace{\gamma^* + \gamma^*} \\
 &  & \phantom{e^+ + e^- + \gamma^*\:}\bentarrow {\rm hadrons} ;
\end{array}
\label{processee}
\eeq
selecting those events in which the incoming leptons produce two 
photons which eventually initiate the hard 
scattering that produces the hadrons. However, it is clear that the 
process in Eq.~(\ref{processee}) is non physical; rather, it has to be 
understood as a shorthand notation for a subset of Feynman diagrams 
contributing to the process that is actually observed,
\beq
e^+ +e^-\,\longrightarrow\,
e^+ +e^- + {\rm hadrons}.
\label{fullproc}
\eeq
Other contributions to the process in Eq.~(\ref{fullproc}) are, for example, 
those in which the incoming $\epem$ pair annihilates into a photon or a $Z$
boson, eventually producing the hadrons and a lepton pair, or those in which 
one (or both) of the two photons in Eq.~(\ref{processee}) is replaced
by a $Z$ boson. However, one can devise a set of cuts such
that the process in Eq.~(\ref{processee}) gives the only non-negligible
contribution to the process in Eq.~(\ref{fullproc}). One can tag 
both of the outgoing leptons, and retain only those events in which the 
scattering angles 
of the leptons are small: in such a way, the contamination due 
to annihilation processes is safely negligible.
Furthermore, small-angle tagging also guarantees that the photon
virtualities are never too large (at LEP2, one typically measures
$Q_i^2={\cal O}(10$~GeV$^2$)); therefore, the contributions from processes
in which a photon is replaced by a $Z$ boson are also negligible.
Thus, it is not difficult to extract the cross section of the 
process \mbox{$\gamma^*\gamma^* \to$~hadrons} from the data relevant 
to the process in Eq.~(\ref{fullproc}).

Our calculation was performed in the massless limit for the final state
quarks. We compared our LO result
to the massless limit of the JAMVG program of Ref.~\cite{Vermaseren:1983cz},
and found perfect agreement.
To study the effect of the NLO corrections, we used the experimental cuts
employed by the L3 Collaboration. The scattered 
electron and positron are required to have energy $E_{1,2}$ larger than 30 
GeV and scattering angle $\theta_{1,2}$ between 30 and 66 mrad. Furthermore, 
the rapidity-like variable $Y$, defined by
\beq
Y=\log\frac{y_1 y_2 S}{\sqrt{Q_1^2 Q_2^2}} ,
\label{YD}
\eeq
is required to lie between 2 and 7 ($y_i$, with $i$~=1, 2, is proportional 
to the
light-cone momentum fraction of the $i^{th}$ virtual photon, and is precisely
defined in Ref.~\cite{Cacciari:2001cb}, 
where a discussion on the properties of $Y$ can also be found). The cross 
sections have been evaluated at $\sqrt{S} = 200$ GeV, including up to
five massless flavours.

We discuss briefly the dependence of our predictions on the
electromagnetic and strong couplings;
our cross sections are $\ord(\aem^4)$ and
we chose to evolve $\aem$ (using one-loop \MS~running)
on an event-by-event basis to the scales set by the virtualities of
the exchanged photons; hence, we replace the Thomson value $\alpha_0
\simeq 1/137$ by $\aem(Q_i^2)$. We treat independently the two photon 
legs: thus, $\alpha_{\rm em}^4$ has to be 
understood as \mbox{$\alpha_{\rm em}^2(Q_1^2)\alpha_{\rm em}^2(Q_2^2)$}.
As for the strong coupling $\as$, we define a default scale $\mu_0$ so 
as to match the order of magnitude of the (inverse of the) interaction 
range,
\beq
\mu_0^2 = \frac{Q_1^2+Q_2^2}{2} +
\left(\frac{k_{1{\sss T}} + k_{2{\sss T}} + k_{3{\sss T}}}{2}\right)^2 \: .
\label{defscale}
\eeq
The renormalization scale $\mu$ entering $\as$ is set 
equal to $\mu_0$ as a default value, and equal to $\mu_0/2$ or $2\mu_0$ when
studying the scale dependence of the cross section. In Eq.~(\ref{defscale}),
the $k_{i{\sss T}}$ are the transverse energies of the outgoing quarks and, 
for three-particle events, the emitted gluon. Since the hard process is 
initiated by the two virtual photons, we study its
properties in the $\gamma^*\gamma^*$ center-of-mass frame. 
We evolve $\as$ to next-to-leading log accuracy, 
with $\as(M_{\sss Z})=0.1181$~\cite{Groom:2000in} (in $\overline{{\rm MS}}$ 
at two loops and with five flavours, this implies 
$\Lambda_{\overline{{\rm MS}}}^{(5)} = 0.2275$ GeV).
In all of the distributions examined~\cite{Cacciari:2001cb}, we found that
the uncertainty related to $\mu$ is always smaller than 
the net effect of including the NLO corrections.
As for the effect of the NLO corrections themselves, we found that, apart
from slightly increasing the cross section with respect to the LO
calculation, they induce visible shape modifications in the $Y$ distribution,
their effect changing from almost nil at the lowest end of the $Y$ spectrum
to a more than 50\% increase at the highest end.

\EPSFIGURE[ht]{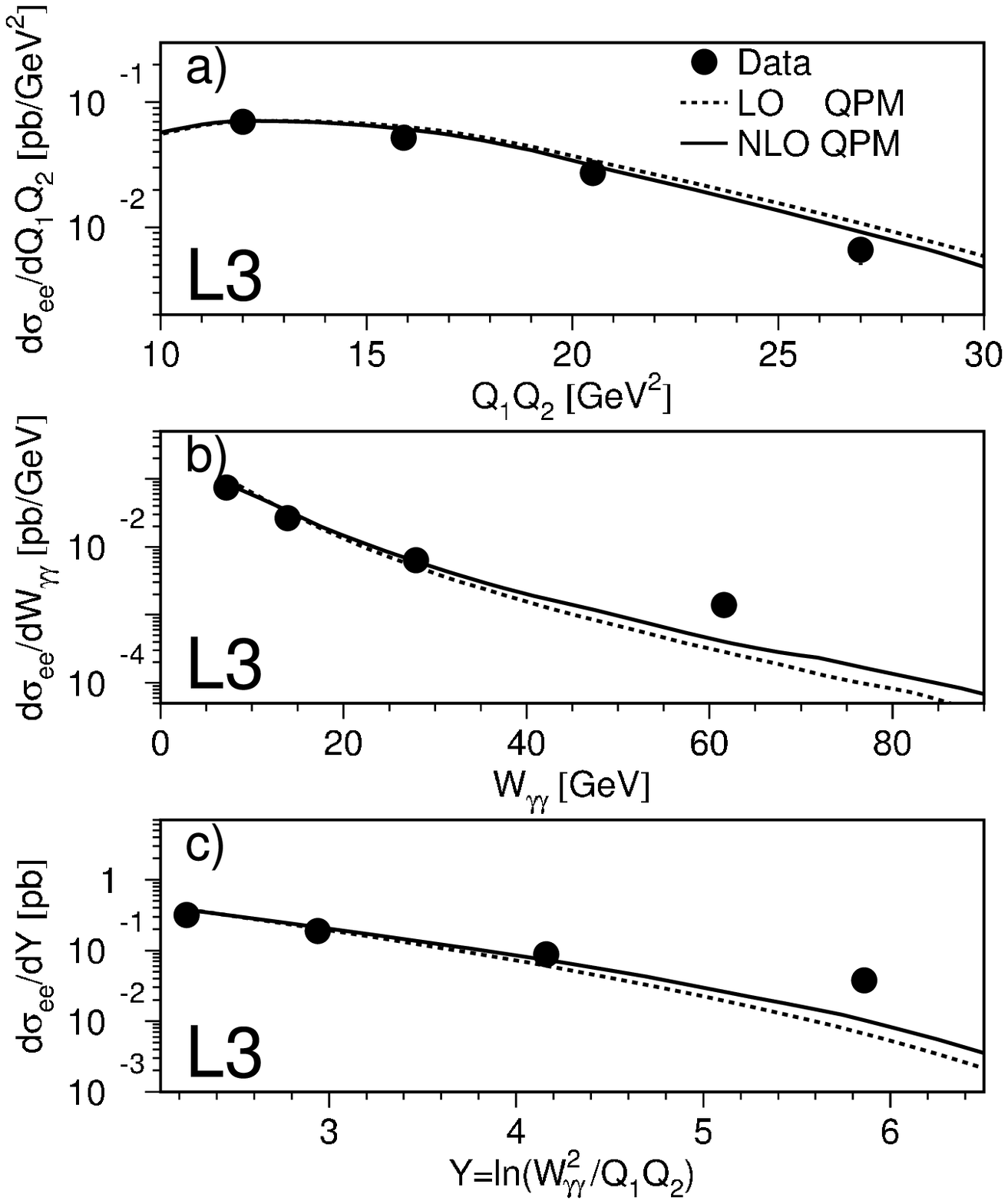,width=0.8\textwidth}{
\label{fig-l3}
Differential cross sections with respect to $Q_1^2 Q_2^2$,
$W_{\gamma\gamma}$ and $Y$ from the L3
Collaboration, compared to leading and next-to-leading order
predictions. The data are taken at $\sqrt{S} = 189 - 202$~GeV. 
The theoretical simulation is always run at $\sqrt{S} = 200$ GeV.}

The L3~\cite{Acciarri:1999ix,lin}, OPAL~\cite{Abbiendi:2001tv} 
and ALEPH~\cite{Prange} Collaborations have
analysed data for hadron production in $\epem$ 
collisions (through $\gamma^*\gamma^*$ scattering) at a center-of-mass 
energy around 200 GeV. L3 made use of the previously mentioned set 
of experimental cuts. Recently L3 re-analysed their data~\cite{lin,L3new}.
The new analysis of L3 is reported in Fig.~\ref{fig-l3}, where the
cross section is presented as a function of the geometric mean of the photon
virtualities $Q_1 Q_2$, the hadron energy $W_{\gamma\gamma}$
and the $Y$ variable, and is compared to our leading and 
next-to-leading order predictions~\cite{Cacciari:2001cb}), 
evaluated at $\sqrt{S} = 200$~GeV.
We note that in the distribution as a function of $Q_1 Q_2$ there is a 
fairly good agreement between theory and data; in the $W_{\gamma\gamma}$
and $Y$ distributions the agreement between theory and data is good at
the low end of the spectrum, but the data tend to 
lie above the theory at the high end of the spectrum\footnote{It is to 
be noted that the scale uncertainties
affecting our predictions are much smaller than the experimental errors.}.
Thus we find a noticeable difference in 
shape between theory and data which, if confirmed, could be interpreted 
as the onset of important higher order effects, perhaps of BFKL type.

We have also studied the effect of the finite mass of the outgoing
heavy quarks in the charm and bottom case, by comparing our results
with the ones obtained with the JAMVG~\cite{Vermaseren:1983cz} code.
Within the L3 set of cuts, such mass effects can be seen to decrease
the LO massless cross section by an amount of the order of  10-15\%.

The data the OPAL Collaboration has 
taken at $\sqrt{S} = 189$ - 202 GeV, making use of
a slightly  different set of cuts are analysed in Ref.~\cite{Abbiendi:2001tv},
and compared there to our NLO predictions~\cite{Cacciari:2001cb}.
For the differential distribution in the variable
$\overline{Y}$ (a variant of $Y$, which tends to it in the high-energy
limit~\cite{Cacciari:2001cb}), a generally good 
agreement within errors can be observed, even though in the largest
$\overline{Y}$ bin the data tend to lie above the prediction.
In the analysis of the data and in the comparison with
our NLO predictions that ALEPH~\cite{Prange} has performed, there
is a good agreement between theory and data in the $W_{\gamma\gamma}$
distribution, while a difference between theory and data is present in
the $Y$ distribution, but only in normalisation and not in shape.
Thus we find in general a good agreement between theory and LEP2 data,
with a discrepancy between theory and L3 data at the highest end 
of the distributions in $W_{\gamma\gamma}$ and $Y$. It would therefore
be of the utmost importance to measure as accurately as possible the 
$Y$ spectrum, in order to perform a precise study of effects beyond NLO 
(such as BFKL dynamics).

\DOUBLEFIGURE[t]{Et-sijet.ps, width=.45\textwidth}
{Deta-Y-jet.ps, width=.45\textwidth}{\label{fig:jetet} 
Transverse energy in single-inclusive jet production.}{\label{fig:jeteta}
Distribution of the rapidity difference between the jets in dijet
production.}

Other production rates of interest are the jet distributions, because
they give us information on different kinematic regions.
We define the jets by means of a $k_{\sss T}$ clustering algorithm,
and set the jet-resolution parameter $D=1$.
In Fig.~\ref{fig:jetet} we show the transverse energy
distribution of single-inclusive jets, considering the cuts $Y>2$ 
and $Y>6$. The first striking feature of this observable is that the 
curves relevant to $Y>2$ and $Y>6$ coincide for $E_{\sss T}>40$~GeV.
In fact, at the threshold (where the jets
are produced at zero rapidity), $W^2=4E_{\sss T}^2$,
we obtain $Y \simeq 6$~\cite{Cacciari:2001cb}. Therefore, the region 
$2<Y<6$ simply does not contribute to events with $E_{\sss T}>40$~GeV:
at $E_{\sss T}=40$~GeV, the two-photon system has just enough
energy, at $Y=6$, to produce the jets. Larger values of $Y$ do
not contribute much, since the $Y$ spectrum is very rapidly
falling at large $Y$'s. When considering
larger transverse momenta, the situation is exactly the same.
We are led to the conclusion that the tail of the 
$E_{\sss T}$ spectrum is dominated by threshold production,
and therefore cannot be reliably predicted by a fixed-order
computation, like ours; a resummation of large threshold 
logarithms is necessary. 
At smaller transverse energies the behaviour of
the radiative corrections displays a pattern similar to that
of total rates. For $Y>2$, NLO and LO results are very close
to each other. For $Y>6$, the radiative
corrections increase sizably the LO result; this is in agreement with
the behaviour of the $Y$ spectrum shown. The
increase is related to the appearance of large logarithms
in the cross section, as it is always the case when two scales 
(here, the small $E_{\sss T}$ and the large hadronic energy) are present.
Next, we argue that the large logarithms in the large-Y region are of 
BFKL type.

In Fig.~\ref{fig:jeteta}, we show the distributions in the
rapidity interval $\Delta\eta$ between the two tagged jets in dijet events, 
for various cuts on $Y$. We select the jets
by imposing a $E_{\sss T}>14$ GeV cut on the transverse energy of the most
energetic jet and requiring $E_{\sss T}>10$ GeV for at least another
jet (in this case, only the NLO results are shown),
in order to avoid the problems that arise in the
case in which such cuts are chosen to be equal.
The most interesting feature of this plot is that it shows that
the large-$Y$ and the large-$\Delta\eta$ regions select the same events:
the distributions relevant to
$Y>2$ (solid line) and to $Y>6$ (dot-dashed line) exactly coincide
for $\Delta\eta>3.5$. This is the same behaviour we observe
in the case of the transverse energy distribution, but the underlying
physics is different. In fact, in this case we also get sizable
contributions away from the threshold; thus, at fixed $E_{\sss T}$,
part of the energy of the two-photon system contributes to the 
longitudinal degrees of freedom, and jets can be produced away from
the central region. 
The large-$Y$ region is thus naturally suitable to study BFKL physics.


\begin{thebibliography}{99}

\bibitem{Acciarri:1999ix}
M.~Acciarri {\it et al.}  [L3 Collaboration],
{\it Measurement of the cross-section for the process gamma* gamma* -->  
hadrons at LEP},
\plb{453}{1999}{333};\\
%%CITATION = PHLTA,B453,333;%%
%\href{\wwwspires?j=PHLTA\%2cB453\%2c333}{SPIRES}

\bibitem{lin}
C.H.~Lin, {\it Double tagged events in two photon collisions at L3
experiment}, Proceedings of the PHOTON 2001 Conf., Ascona.

%\cite{Abbiendi:2001tv}
\bibitem{Abbiendi:2001tv}
G.~Abbiendi {\it et al.}  [OPAL Collaboration],
{\it Measurement of the hadronic cross-section for the scattering of two  
virtual photons at LEP}, \hepex{0110006}.
%%CITATION = HEP-EX 0110006;%%

\bibitem{Prange}
G.~Prange, {\it Measurement of the hadronic cross section of double tagged
$\gamma\gamma$ events at ALEPH}, Proceedings of the PHOTON 2001 Conf., Ascona.

\bibitem{Bartels:1996ke}
J.~Bartels, A.~De Roeck and H.~Lotter,
{\it The gamma* gamma* total cross section and the BFKL pomeron 
at e+ e-  colliders},
\plb{389}{1996}{742}
[\hepph{9608401}];\\
%%CITATION = HEP-PH 9608401;%%
%\href{\wwwspires?eprint=HEP-PH/9608401}{SPIRES}
S.~J.~Brodsky, F.~Hautmann and D.~E.~Soper,
{\it Virtual photon scattering at high energies as a probe of the short  
distance pomeron},
\prd{56}{1997}{6957} [\hepph{9706427}].
%%CITATION = HEP-PH 9706427;%%
%\href{\wwwspires?eprint=HEP-PH/9706427}{SPIRES}

%\cite{Cacciari:2001cb}
\bibitem{Cacciari:2001cb}
M.~Cacciari, V.~Del Duca, S.~Frixione and Z.~Tr\'ocs\'anyi,
{\it QCD radiative corrections to gamma* gamma* $\to$ hadrons},
\jhep{02}{2001}{029} [\hepph{0011368}].
%%CITATION = HEP-PH 0011368;%%

\bibitem{Vermaseren:1983cz}
J.~A.~Vermaseren,
{\it Two Photon Processes At Very High-Energies},
\npb{229}{1983}{347}.
%%CITATION = NUPHA,B229,347;%%
%\href{\wwwspires?j=NUPHA\%2cB229\%2c347}{SPIRES}

\bibitem{Groom:2000in}
D.~E.~Groom {\it et al.},
%``Review of particle physics,''
\epjc{15}{2000}{1}.
%%CITATION = EPHJA,C15,1;%%

\bibitem{L3new}
M.~Przybycien in Proceedings of the EPS International Conference on
High Energy Physics, Budapest, 2001 (D. Horvath, P. Levai, A. Patkos,
eds.), JHEP (http://jhep.sissa.it/) Proceedings Section,
PrHEP-hep2001/044.

\end{thebibliography}
\end{document}